# COORDINATE DESCENT ALGORITHMS FOR LASSO PENALIZED REGRESSION[1]

BY TONG TONG WU AND KENNETH LANGE

*University of Maryland, College Park and University of California, Los Angeles*

Imposition of a lasso penalty shrinks parameter estimates toward zero and performs continuous model selection. Lasso penalized regression is capable of handling linear regression problems where the number of predictors far exceeds the number of cases. This paper tests two exceptionally fast algorithms for estimating regression coefficients with a lasso penalty. The previously known $\ell_2$ algorithm is based on cyclic coordinate descent. Our new $\ell_1$ algorithm is based on greedy coordinate descent and Edgeworth's algorithm for ordinary $\ell_1$ regression. Each algorithm relies on a tuning constant that can be chosen by cross-validation. In some regression problems it is natural to group parameters and penalize parameters group by group rather than separately. If the group penalty is proportional to the Euclidean norm of the parameters of the group, then it is possible to majorize the norm and reduce parameter estimation to $\ell_2$ regression with a lasso penalty. Thus, the existing algorithm can be extended to novel settings. Each of the algorithms discussed is tested via either simulated or real data or both. The Appendix proves that a greedy form of the $\ell_2$ algorithm converges to the minimum value of the objective function.

**1. Introduction.** This paper explores fast algorithms for lasso penalized regression [Chen et al. (1998), Claerbout and Muir (1973), Santosa and Symes (1986), Taylor et al. (1979) and Tibshirani (1996)]. The lasso performs continuous model selection and enforces sparse solutions in problems where the number of predictors $p$ exceeds the number of cases $n$. In the regression setting, let $y_i$ be the response for case $i$, $x_{ij}$ be the value of predictor $j$ for case $i$, and $\beta_j$ be the regression coefficient corresponding to predictor $j$. The intercept $\mu$ is ignored in the lasso penalty, whose strength is determined by the

Received May 2007; revised October 2007.
[1]Supported in part by NIH Grants GM53275 and MH59490.
*Key words and phrases.* Model selection, Edgeworth's algorithm, cyclic, greedy, consistency, convergence.

positive tuning constant $\lambda$. Given the parameter vector $\theta = (\mu, \beta_1, \ldots, \beta_p)^t$ and the loss function $g(\theta)$, lasso penalized regression can be phrased as minimizing the criterion

$$(1) \qquad f(\theta) = g(\theta) + \lambda \sum_{j=1}^{p} |\beta_j|,$$

where $g(\theta)$ equals $\sum_{i=1}^{n} |y_i - \mu - \sum_{j=1}^{p} x_{ij}\beta_j|$ for $\ell_1$ regression and $g(\theta)$ equals $\frac{1}{2}\sum_{i=1}^{n}(y_i - \mu - \sum_{j=1}^{p} x_{ij}\beta_j)^2$ for $\ell_2$ regression.

The lasso penalty $\lambda \sum_{j=1}^{p} |\beta_j|$ shrinks each $\beta_j$ toward the origin and tends to discourage models with large numbers of marginally relevant predictors. The lasso penalty is more effective in deleting irrelevant predictors than a ridge penalty $\lambda \sum_{j=1}^{p} \beta_j^2$ because $|b|$ is much bigger than $b^2$ for small $b$. When protection against outliers is a major concern, $\ell_1$ regression is preferable to $\ell_2$ regression [Wang et al. (2006a)].

Lasso penalized estimation raises two issues. First, what is the most effective method of minimizing the objective function (1)? Second, how does one choose the tuning parameter $\lambda$? Although the natural answer to the second question is cross-validation, the issue of efficient computation arises here as well. We will discuss a useful approach in Section 6. The answer to the first question is less obvious. Standard methods of regression involve matrix diagonalization, matrix inversion, or, at the very least, the solution of large systems of linear equations. Because the number of arithmetic operations for these processes scales as the cube of the number of predictors, problems with thousands of predictors appear intractable. Recent research has shown this assessment to be too pessimistic [Candes and Tao (2007), Park and Hastie (2006a, 2006b) and Wang et al. (2006a)]. In the current paper we highlight the method of coordinate descent. Our reasons for liking coordinate descent boil down to simplicity, speed and stability.

Fu (1998) and Daubechies et al. (2004) explicitly suggest coordinate descent for lasso penalized $\ell_2$ regression. For inexplicable reasons, they did not follow up their theoretical suggestions with numerical confirmation for highly underdetermined problems. Claerbout and Muir (1973) note that lasso penalized $\ell_1$ regression also yields to coordinate descent. Both methods are incredibly quick and have the potential to revolutionize data mining. The competing linear programming algorithm of Wang et al. (2006a) for penalized $\ell_1$ regression is motivated by the problem of choosing the tuning parameter $\lambda$. Their algorithm follows the central path determined by the minimum of $f(\theta)$ as a function of $\lambda$. This procedure reveals exactly when each estimated $\beta_j$ enters the linear prediction model. The central path method is also applicable to penalized $\ell_2$ regression and penalized estimation with generalized linear models [Park and Hastie (2006b)].

Besides introducing a modification of the $\ell_1$ coordinate descent algorithm, we want to comment on group selection in $\ell_2$ regression. To set the stage for both purposes, we will review the previous work of Fu (1998) and Daubechies et al. (2004). We approach $\ell_1$ regression through the nearly forgotten algorithm of Edgeworth (1887, 1888), which for a long time was considered a competitor of least squares. Portnoy and Koenker (1997) trace the history of the algorithm from Boscovich to Laplace to Edgeworth. It is fair to say that the algorithm has managed to cling to life despite decades of obscurity both before and after its rediscovery by Edgeworth. Armstrong and Kung (1978) published a computer implementation of Edgeworth's algorithm in *Applied Statistics*. Unfortunately, this version is limited to simple linear regression. We adapt the Claerbout and Muir (1973) version of Edgeworth's algorithm to perform greedy coordinate descent. The resulting $\ell_1$ algorithm is faster than cyclic coordinate descent in $\ell_2$ regression.

Many data sets involve groups of correlated predictors. For example, in gene microarray experiments, genes can sometimes be grouped into biochemical pathways subject to genetic coregulation. Expression levels for genes in the same pathway are expected to be highly correlated. In such situations it is prudent to group genes and design penalties that apply to entire groups. Several authors have taken up the challenge of penalized estimation in this context [Zou and Hastie (2005), Yuan and Lin (2006) and Zhao et al. (2006)]. In the current paper we will demonstrate that cyclic coordinate descent is compatible with penalties constructed from the Euclidean norms of parameter groups. We attack penalized estimation by combining cyclic coordinate descent with penalty majorization. This replaces the nonquadratic norm penalties by $\ell_1$ or $\ell_2$ penalties. The resulting algorithm is reminiscent of the generic MM algorithm for parameter estimation [Lange (2004)].

In the remainder of the paper Section 2 reviews cyclic coordinate descent for penalized $\ell_2$ regression, and Section 3 develops Edgeworth's algorithm for penalized $\ell_1$ regression. Section 4 briefly discusses convergence of coordinate descent in penalized $\ell_2$ regression; the Appendix proves convergence for greedy coordinate descent. Section 5 amends the $\ell_2$ algorithm to take into account grouped parameters, and Section 6 gives some guidance on how to select tuning constants. Sections 7 and 8 test the algorithms on simulated and real data, and Section 9 summarizes their strengths and suggests new avenues of research.

Finally, we would like to draw the reader's attention to the recent paper of Friedman et al. (2007) in this journal on coordinate descent and the fused lasso. Their paper has substantial overlap and substantial differences with ours. The two papers were written independently and concurrently.

**2. Cyclic coordinate descent for $\ell_2$ regression.** Coordinate descent comes in several varieties. The standard version cycles through the parameters and updates each in turn. An alternative version is greedy and updates the parameter giving the largest decrease in the objective function. Because it is impossible to tell in advance which parameter is best to update, the greedy version uses the surrogate criterion of steepest descent. In other words, for each parameter we compute forward and backward directional derivatives and then update the parameter with the most negative directional derivative, either forward or backward. The overhead of keeping track of these directional derivatives works to the detriment of the greedy method. For $\ell_1$ regression, the overhead is relatively light, and greedy coordinate descent is substantially faster than cyclic coordinate descent.

Although the lasso penalty is nondifferentiable, it does possess directional derivatives along each forward or backward coordinate direction. For instance, if $e_k$ is the coordinate direction along which $\beta_k$ varies, then the objective function (1) has directional derivatives

$$d_{e_k} f(\theta) = \lim_{\tau \downarrow 0} \frac{f(\theta + \tau e_k) - f(\theta)}{\tau} = d_{e_k} g(\theta) + \begin{cases} \lambda, & \beta_k \geq 0, \\ -\lambda, & \beta_k < 0, \end{cases}$$

and

$$d_{-e_k} f(\theta) = \lim_{\tau \downarrow 0} \frac{f(\theta - \tau e_k) - f(\theta)}{\tau} = d_{-e_k} g(\theta) + \begin{cases} -\lambda, & \beta_k > 0, \\ \lambda, & \beta_k \leq 0. \end{cases}$$

In $\ell_2$ regression, the function $g(\theta)$ is differentiable. Therefore, its directional derivative along $e_k$ coincides with its ordinary partial derivative

$$\frac{\partial}{\partial \beta_k} g(\theta) = -\sum_{i=1}^{n} \left( y_i - \mu - \sum_{j=1}^{p} x_{ij} \beta_j \right) x_{ik},$$

and its directional derivative along $-e_k$ coincides with the negative of its ordinary partial derivative. In $\ell_1$ regression, the coordinate direction derivatives are

$$d_{e_k} g(\theta) = \sum_{i=1}^{n} \begin{cases} -x_{ik}, & y_i - \mu - x_i^t \beta > 0, \\ x_{ik}, & y_i - \mu - x_i^t \beta < 0, \\ |x_{ik}|, & y_i - \mu - x_i^t \beta = 0, \end{cases}$$

and

$$d_{-e_k} g(\theta) = \sum_{i=1}^{n} \begin{cases} x_{ik}, & y_i - \mu - x_i^t \beta > 0, \\ -x_{ik}, & y_i - \mu - x_i^t \beta < 0, \\ |x_{ik}|, & y_i - \mu - x_i^t \beta = 0, \end{cases}$$

where $x_i^t$ is the row vector $(x_{i1}, \ldots, x_{ip})$.

In cyclic coordinate descent we evaluate $d_{e_k} f(\theta)$ and $d_{-e_k} f(\theta)$. If both are nonnegative, then we skip the update for $\beta_k$. This decision is defensible when $g(\theta)$ is convex because the sign of a directional derivative fully determines whether improvement can be made in that direction. If either directional derivative is negative, then we must solve for the minimum in that direction. Because the objective function $f(\theta)$ is convex, it is impossible for both directional derivatives $d_{e_k}f(\theta)$ and $d_{-e_k}f(\theta)$ to be negative.

In underdetermined problems with just a few relevant predictors, most updates are skipped, and the corresponding parameters never budge from their starting values of 0. This simple fact plus the complete absence of matrix operations explains the speed of cyclic coordinate descent. It inherits its numerical stability from the descent property of each update.

Fu (1998) derived cyclic coordinate descent algorithms for $\ell_2$ regression with penalties $\lambda \sum_j |\beta_j|^\alpha$ with $\alpha \geq 1$. With a lasso penalty ($\alpha = 1$), the update of the intercept parameter can be written as

$$\hat{\mu} = \frac{1}{n}\sum_{i=1}^{n}(y_i - x_i^t\beta) = \mu - \frac{1}{n}\frac{\partial}{\partial \mu}g(\theta).$$

For the parameter $\beta_k$, there are separate solutions to the left and right of 0. These amount to

$$\hat{\beta}_{k,-} = \min\left\{0, \beta_k - \frac{\frac{\partial}{\partial \beta_k}g(\theta) - \lambda}{\sum_{i=1}^{n} x_{ik}^2}\right\},$$

$$\hat{\beta}_{k,+} = \max\left\{0, \beta_k - \frac{\frac{\partial}{\partial \beta_k}g(\theta) + \lambda}{\sum_{i=1}^{n} x_{ik}^2}\right\}.$$

Only one of these two solutions can be nonzero. The partial derivatives

$$\frac{\partial}{\partial \mu}g(\theta) = -\sum_{i=1}^{n} r_i, \qquad \frac{\partial}{\partial \beta_k}g(\theta) = -\sum_{i=1}^{n} r_i x_{ik}$$

of $g(\theta)$ are easy to compute if we keep track of the residuals $r_i = y_i - \mu - x_i^t\beta$. The residual $r_i$ is reset to $r_i + \mu - \hat{\mu}$ when $\mu$ is updated and to $r_i + x_{ik}(\beta_k - \hat{\beta}_k)$ when $\beta_k$ is updated. Organizing all updates around residuals promotes fast evaluation of $g(\theta)$.

**3. Greedy coordinate descent for $\ell_1$ regression.** In greedy coordinate descent, we update the parameter $\theta_k$ giving the most negative value of $\min\{df_{e_k}(\theta), df_{-e_k}(\theta)\}$. If none of the coordinate directional derivatives is negative, then no further progress can be made. In lasso constrained $\ell_1$ regression greedy coordinate descent is quick because directional derivatives are trivial to update. Indeed, if updating $\beta_k$ does not alter the sign of the residual $r_i = y_i - \mu - x_i^t\beta$ for case $i$, then the contributions of case $i$ to the various directional derivatives do not change. When the residual $r_i$ becomes 0 or changes sign, these contributions are modified by simply adding or subtracting entries of the design matrix. Similar considerations apply when $\mu$ is updated.

To illustrate Edgeworth's algorithm in action, consider minimizing the two-parameter model $g(\theta) = \sum_{i=1}^{n} |y_i - \mu - x_i\beta|$ with a single slope $\beta$. To update $\mu$, we recall the well-known connection between $\ell_1$ regression and medians and replace $\mu$ for fixed $\beta$ by the sample median of the numbers $z_i = y_i - x_i\beta$. This action drives $g(\theta)$ downhill. Updating $\beta$ for $\mu$ fixed depends on writing

$$g(\theta) = \sum_{i=1}^{n} |x_i| \left| \frac{y_i - \mu}{x_i} - \beta \right|,$$

sorting the numbers $z_i = (y_i - \mu)/x_i$, and finding the weighted median with weight $w_i = |x_i|$ assigned to $z_i$. We replace $\beta$ by the order statistic $z_{[i]}$ whose index $i$ satisfies

$$\sum_{j=1}^{i-1} w_{[j]} < \tfrac{1}{2} \sum_{j=1}^{n} w_{[j]}, \qquad \sum_{j=1}^{i} w_{[j]} \geq \tfrac{1}{2} \sum_{j=1}^{n} w_{[j]}.$$

With more than a single predictor, we update parameter $\beta_k$ by writing

$$g(\theta) = \sum_{i=1}^{n} |x_{ik}| \left| \frac{y_i - \mu - \sum_{j \neq k} x_{ij}\beta_j}{x_{ik}} - \beta_k \right|,$$

and finding the weighted median.

Two criticisms have been leveled at Edgeworth's algorithm. First, although it drives the objective function steadily downhill, it sometimes converges to an inferior point. Li and Arce (2004) give an example involving the data values $(0.3, -1.0)$, $(-0.4, -0.1)$, $(-2.0, -2.9)$, $(-0.9, -2.4)$ and $(-1.1, 2.2)$ for the pairs $(x_i, y_i)$ and parameter starting values $(\mu, \beta) = (3.5, -1.0)$. Unfortunately, Li and Arce's suggested improvement to Edgeworth's algorithm does not generalize readily to multivariate linear regression. The second criticism is that convergence often occurs in a slow seesaw pattern. These defects are not fatal.

In fact, our numerical examples show that the greedy version of Edgeworth's algorithm performs well on most practical problems. It has little difficulty in picking out relevant predictors, and it usually takes less computing time to converge than $\ell_2$ regression by cyclic coordinate descent. In $\ell_1$ regression, greedy coordinate descent is considerably faster than cyclic coordinate descent, probably because greedy coordinate descent attacks the significant predictors early on before it gets trapped by an inferior point.

Implementing Edgeworth's algorithm with a lasso penalty requires viewing the penalty terms as the absolute values of pseudo-residuals. Thus, we write $\lambda|\beta_j| = |y - x^t\theta|$ by taking $y = 0$ and $x_k = \lambda 1_{\{k=j\}}$. Edgeworth's algorithm now applies.

Because the $\ell_1$ objective function is nondifferentiable, it is difficult to understand the theoretical properties of $\ell_1$ estimators. Our supplementary appendix [Wu and Lange (2008)] demonstrates the weak consistency of penalized $\ell_1$ estimators. The proof there builds on the previous work of Oberhofer (1983) on nonlinear $\ell_1$ regression. Since we only consider linear models, it is possible to relax and clarify his stated regularity conditions. Zhao and Yu (2006) summarize and extend previous consistency results for $\ell_2$ penalized estimators.

**4. Convergence of the algorithms.** The counterexample cited for Edgeworth's algorithm shows that it may not converge to a minimum point. The question of convergence for the $\ell_2$ algorithms is more interesting. Textbook treatments of convergence for cyclic coordinate descent are predicated on the assumption that the objective function $f(\theta)$ is continuously differentiable. For example, see Proposition 5.32 of Ruszczyński (2006). Coordinate descent may fail for a nondifferentiable function because all directional derivatives must be nonnegative at a minimum point. It does not suffice for just the directional derivatives along the coordinate directions to be nonnegative. Unfortunately, the lasso penalty is nondifferentiable. The more general theory of Tseng (2001) does cover cyclic coordinate descent in $\ell_2$ regression, but it does not apply to greedy coordinate descent. In the Appendix we demonstrate the following proposition.

PROPOSITION 1. *Every cluster point of the $\ell_2$ greedy coordinate descent algorithm is a minimum point of the objective function $f(\theta)$. If the minimum point is unique, then the algorithm converges to it. If the algorithm converges, then its limit is a minimum point.*

Our qualitative theory does not specify the rate of convergence. Readers may want to compare our treatment of convergence to the treatment of Fu (1998).

It would also be helpful to identify a simple sufficient condition making the minimum point unique. Ordinarily, uniqueness is proved by establishing the strict convexity of the objective function. If the problem is overdetermined or the penalty is a ridge penalty, then this is an easy task. For underdetermined problems with lasso penalties, strict convexity can fail. Of course, strict convexity is not necessary for a unique minimum; linear programming is full of examples to the contrary. Based on a conversation with Emanuel Candes, we conjecture that almost all (with respect to Lebesgue measure) design matrices lead to a unique minimum.

**5. $\ell_2$ regression with group penalties.** The issues of modeling and fast estimation are also intertwined with grouped effects, where we want coordinated penalties that tend to include or exclude all of the parameters in a group. Suppose that the $\beta$ parameters occur in $q$ disjoint groups and $\gamma_j$ denotes the parameter vector for group $j$. The lasso penalty $\lambda \|\gamma_j\|_1$ separates parameters and does not qualify as a sensible group penalty. For the same reason the scaled sum of squares $\lambda \|\gamma_j\|_2^2$ is disqualified. However, the scaled Euclidean norm $\lambda \|\gamma_j\|_2$ is an ideal group penalty. It couples the parameters, it preserves convexity, and, as we show in a moment, it meshes well with cyclic coordinate descent in $\ell_2$ regression.

To understand its grouping tendency, note that the directional derivative of $\|\gamma_j\|_2$ along $e_{jk}$, the coordinate vector corresponding to $\gamma_{jk}$, is 1 when $\gamma_j = \mathbf{0}$ and is 0 when $\gamma_j \neq \mathbf{0}$ and $\gamma_{jk} = 0$. Thus, if any parameter $\gamma_{jl}$, $l \neq k$, is nonzero, it becomes easier for $\gamma_{jk}$ to move away from 0. Recall that for a parameter to move away from 0, the forward or backward directional derivative of the objective function must be negative. If a penalty contribution to these directional derivatives drops from 1 to 0, then the directional derivatives are more likely to be negative.

In $\ell_2$ regression with grouping effects, we recommend minimizing the objective function

$$f(\theta) = g(\theta) + \lambda_2 \sum_{j=1}^{q} \|\gamma_j\|_2 + \lambda_1 \sum_{j=1}^{q} \|\gamma_j\|_1,$$

where $g(\theta)$ is the residual sum of squares. If the tuning parameter $\lambda_2 = 0$, then the penalty reduces to the lasso. On the one hand when $\lambda_1 = 0$, only group penalties enter the picture. The mixed penalties with $\lambda_1 > 0$ and $\lambda_2 > 0$ enforce shrinkage in both ways. All mixed penalties are norms and therefore convex functions. Nonconvex penalties complicate optimization and should be avoided whenever possible.

At each stage of cyclic coordinate descent, we are required to minimize $g(\theta) + \lambda_2 \|\gamma_j\|_2 + \lambda_1 \|\gamma_j\|_1$ with respect to a component $\gamma_{jk}$ of some $\gamma_j$. If $\gamma_j = \mathbf{0}$, then $\|\gamma_j\|_2 = |\gamma_{jk}|$ as a function of $\gamma_{jk}$. Thus, minimization with respect to $\gamma_{jk}$ reduces to the standard update for $\ell_2$ regression with a lasso penalty. The lasso tuning parameter $\lambda$ equals $\lambda_1 + \lambda_2$ is this situation. When $\gamma_j \neq \mathbf{0}$, the standard update does not apply.

However, there is an alternative update that stays within the framework of penalized $\ell_2$ regression. This alternative involves majorizing the objective function and is motivated by the MM algorithm for parameter estimation [Lange (2004)]. In view of the concavity of the square root function $\sqrt{t}$, we have the inequality

(2) $$\|\gamma_j\|_2 \leq \|\gamma_j^m\|_2 + \frac{1}{2\|\gamma_j^m\|_2}(\|\gamma_j\|_2^2 - \|\gamma_j^m\|_2^2),$$

where the superscript $m$ indicates iteration number. Equality prevails whenever $\gamma_j = \gamma_j^m$. The right-hand side of inequality (2) is said to majorize the left-hand side. This simple majorization leads to the additional majorization

$$g(\theta) + \lambda_2 \|\gamma_j\|_2 + \lambda_1 \|\gamma_j\|_1$$
$$\leq g(\theta) + \lambda_2 \left[ \|\gamma_j^m\|_2 + \frac{1}{2\|\gamma_j^m\|_2} (\|\gamma_j\|_2^2 - \|\gamma_j^m\|_2^2) \right] + \lambda_1 \|\gamma_j\|_1.$$

As a function of $\gamma_{jk}$ ignoring $\gamma_{jk}^m$, the second majorization amounts to a quadratic plus a lasso penalty. Fortunately, we know how to minimize such a surrogate function. According to the arguments justifying the descent property of the MM algorithm, minimizing the surrogate is guaranteed to drive the objective function downhill.

To summarize, grouped effects can be handled by introducing penalties defined by the Euclidean norms of the grouped parameters. Updating a parameter follows the standard recipe when the other parameters of its group are fixed at 0. If one of the other parameters from its group is nonzero, then we majorize the objective function and minimize the surrogate function with respect to the designated parameter. Again, the update relies on the standard recipe. Although convergence may be slowed by majorization, it is consistent with cyclic coordinate descent and preserves the structure of the updates.

**6. Selection of the tuning constant $\lambda$.** As we mentioned earlier, selection of the tuning constant $\lambda$ can be guided by cross-validation. This is a one-dimensional problem, so inspection of the graph of the cross-validation error curve $c(\lambda)$ suffices in a practical sense. Recall that in $k$-fold cross-validation, one divides the data into $k$ equal batches (subsamples) and estimates parameters $k$ times, leaving one batch out per time. The testing error for each omitted batch is computed using the estimates derived from the remaining batches, and $c(\lambda)$ is computed by averaging testing error across the $k$ batches. In principle, one can substitute other criterion for average cross-validation error. For instance, we could define $c(\lambda)$ by AIC or BIC criteria. For the sake of brevity, we will rest content with cross-validation error.

Evaluating $c(\lambda)$ on a grid of points can be computationally inefficient, particularly if grid points occur near $\lambda = 0$. Although we recommend grid sampling on important problems, it is useful to pursue shortcuts. One shortcut combines bracketing and golden section search. Because coordinate descent is fastest when $\lambda$ is large and the vast majority of $\beta_j$ are estimated as 0, it makes sense to start looking for a bracketing triple with a very large value $\lambda_0$ and work downward. One then repeatedly reduces $\lambda$ by a fixed proportion $r \in (0,1)$ until the condition $c(\lambda_{k+1}) > c(\lambda_k)$ first occurs. This quickly identifies a bracketing triple $\lambda_{k-1} > \lambda_k > \lambda_{k+1}$ with $\lambda_k = r^k \lambda_0$ giving the smallest value of $c(\lambda)$. One can now apply golden section search to minimize $c(\lambda)$ on the interval $(\lambda_{k+1}, \lambda_{k-1})$. With grouped parameters, finding the best pair of tuning parameters $(\lambda_1, \lambda_2)$ is considerably more difficult. As a rough guess, we recommend consideration of the three cases: (a) $\lambda_1 = 0$, (b) $\lambda_2 = 0$ and (c) $\lambda_1 = \lambda_2$. These one-dimensional slices yield to bracketing and golden section search.

Selection of the tuning constant $\lambda$ has implications in setting the initial value of $\theta$. For a single $\lambda$, we recommend setting $\theta^0 = \mathbf{0}$ and all residuals $r_i = 0$. As $\lambda$ decreases, we expect current predictors to be retained and possibly new ones to enter. If we estimate $\hat{\theta}$ for a given $\lambda$, then it makes sense to start with $\hat{\theta}$ and the corresponding residuals for a nearby but smaller value of $\lambda$. This tactic builds on already good estimates, reduces the number of iterations until convergence and saves considerable time overall in evaluating the $c(\lambda)$ curve.

**7. Analysis of simulated data.** In evaluating the performance of the coordinate descent methods, we put special emphasis on the underdetermined setting $p \gg n$ highlighted by Wang et al. (2006a). In the regression model

$$y_i = \mu + \sum_{j=1}^{p} x_{ij}\beta_j + \epsilon_i,$$

we assume that the random errors $\epsilon_i$ are independent and follow either a standard normal distribution or a Laplace (double exponential) distribution with scale 1. The predictor vectors $x_i$ represent a random sample from a multivariate normal distribution whose marginals are standard normal and whose pairwise correlations are

$$\mathrm{Cov}(X_{ij}, X_{ik}) = \begin{cases} \rho, & j \leq 10 \text{ and } k \leq 10, \\ 0, & \text{otherwise.} \end{cases}$$

In every simulation the true parameter values are $\beta_j = 1$ for $1 \leq j \leq 5$ and $\beta_j = 0$ for $j > 5$.

The quality of the parameter estimates and the optimal value of $\lambda$ are naturally of interest. To ameliorate the shrinkage of nonzero estimates for a particular $\lambda$, we always re-estimate the active parameters, omitting the inactive parameters and the lasso penalty. This yields better parameter estimates for testing purposes. The choice of $\lambda$ depends on the 10-fold average cross-validation error curve $c(\lambda)$. We sample $c(\lambda)$ on a grid and find its minimum by bracketing and golden section search as previously described. Given the optimal $\lambda$, we re-estimate parameters from the data as a whole and compute prediction error on a testing data set of 20,000 additional cases. It is instructive to compare this approximate prediction error to the true prediction error using the true regression coefficients.

Table 1 reports average prediction errors in $\ell_1$ regression based on 50 replicates and problem sizes of $(p, n) = (5000, 200)$, $(p, n) = (5000, 500)$ and

COORDINATE DESCENT ALGORITHMS FOR PENALIZED REGRESSION 11$(p, n) = (50000, 500)$ and both independent predictors ($\rho = 0$) and highly correlated predictors ($\rho = 0.8$). The average number of predictors selected is listed as $N_{\text{nonzero}}$, and the average number of true predictors selected is listed as $N_{\text{true}}$. Average computing time in seconds at the optimal value of $\lambda$ is recorded in the last column of the table. On our personal computers, computing times are remarkably fast even for data sets as large as $(p, n) = (50000, 500)$. It is clear that the coordinate descent algorithms keep almost all true predictors while discarding the vast majority of irrelevant ones. Approximate prediction errors are very close to true prediction errors.

To better understand the impact of the tuning constant $\lambda$ in penalized $\ell_1$ regression, we plot prediction error versus $\lambda$ in the left panel of Figure 1 for one realization of the data. Here we take $(p, n) = (5000, 200)$, independent predictors, and Laplace errors. The solid line shows prediction errors based on 10-fold cross-validation. The dashed line shows prediction errors based on the 20,000 testing cases. It is noteworthy that cross-validation underestimates the optimal value of $\lambda$ suggested by testing error. The optimal $\lambda$ based on 10-fold cross-validation is around 45, and the optimal $\lambda$ based on 20,000 testing cases is around 50. Many statisticians are comfortable with this conservative bias of cross-validation.

TABLE 1
*Simulation results for $\ell_1$ regression with a lasso penalty. Standard errors of estimates appear in parentheses. The left error column is testing error under the true parameter values; the right error column is testing error under the estimated parameter values*

| | | | | $\beta = (1, 1, 1, 1, 1, 0, \ldots, 0)$ | | | | |
|---|---|---|---|---|---|---|---|---|
| **Distribution** | $(p, n)$ | $\rho$ | **Error** | $\lambda$ | **Error** | $N_{\text{nonzero}}$ | $N_{\text{true}}$ | **Time** |
| Laplace | (5000, 200) | 0.00 | 0.99 | 44.05 | 1.11 | 14.06 | 5.00 | 0.02 |
| | | | | (4.43) | (0.08) | (8.63) | (0.00) | (0.01) |
| Laplace | (5000, 200) | 0.80 | 0.99 | 72.39 | 1.04 | 6.80 | 5.00 | 0.04 |
| | | | | (13.60) | (0.03) | (1.57) | (0.00) | (0.01) |
| Laplace | (5000, 500) | 0.00 | 1.01 | 101.51 | 1.01 | 5.18 | 5.00 | 0.09 |
| | | | | (10.58) | (0.01) | (0.65) | (0.00) | (0.01) |
| Laplace | (5000, 500) | 0.80 | 1.01 | 132.88 | 1.01 | 6.22 | 5.00 | 0.09 |
| | | | | (34.93) | (0.01) | (1.27) | (0.00) | (0.02) |
| Laplace | (50000, 500) | 0.00 | 1.01 | 109.21 | 1.01 | 5.12 | 5.00 | 0.36 |
| | | | | (9.45) | (0.01) | (0.32) | (0.00) | (0.04) |
| Laplace | (50000, 500) | 0.80 | 1.01 | 150.44 | 1.01 | 6.44 | 5.00 | 1.59 |
| | | | | (43.68) | (0.01) | (1.12) | (0.00) | (0.33) |
| Normal | (5000, 200) | 0.00 | 0.80 | 48.46 | 0.84 | 7.84 | 5.00 | 0.03 |
| | | | | (3.58) | (0.04) | (3.31) | (0.00) | (0.02) |
| Normal | (5000, 200) | 0.80 | 0.80 | 76.71 | 0.82 | 6.08 | 4.98 | 0.04 |
| | | | | (14.44) | (0.02) | (0.93) | (0.14) | (0.01) |
| Normal | (5000, 500) | 0.00 | 0.80 | 101.71 | 0.81 | 5.48 | 5.00 | 0.05 |
| | | | | (16.98) | (0.01) | (1.19) | (0.00) | (0.01) |
| Normal | (5000, 500) | 0.80 | 0.80 | 150.83 | 0.81 | 6.04 | 5.00 | 0.10 |
| | | | | (47.28) | (0.01) | (1.06) | (0.00) | (0.03) |
| Normal | (50000, 500) | 0.00 | 0.80 | 112.03 | 0.81 | 5.10 | 5.00 | 0.75 |
| | | | | (12.03) | (0.01) | (0.46) | (0.00) | (0.07) |
| Normal | (50000, 500) | 0.80 | 0.80 | 149.83 | 0.81 | 6.36 | 5.00 | 2.02 |
| | | | | (42.17) | (0.01) | (1.09) | (0.00) | (0.45) |

Table 2 reports the results of $\ell_2$ regression under the same conditions except for normal errors. The cyclic coordinate descent algorithm for $\ell_2$ regression is slightly more reliable, slightly less parsimonious and considerably slower than the greedy coordinate descent algorithm for $\ell_1$ regression. The right panel of Figure 1 plots cross-validation error and approximate prediction error versus $\lambda$.

Table 3 compares the speed and performance of three algorithms for lasso penalized $\ell_2$ regression on one realization of each of the simulated data sets with normally distributed errors. Because LARS [Hastie and Efron (2007)] is considered by many to be the best competing algorithm, it is reasonable to limit our comparison of the two versions of coordinate descent to LARS. Three conclusions can be drawn from Table 3. First, cyclic coordinate descent is definitely faster than greedy coordinate descent for $\ell_2$ regression. Second, both methods are considerably faster and more robust than LARS. Third, both methods are more successful than LARS in model selection. Note that the error estimates in the table for the coordinate descent algorithms reflect the re-estimation step mentioned earlier. This may put LARS at a disadvantage.

Table 4 compares the greedy coordinate descent and cyclic coordinate descent algorithms for $\ell_1$ regression. The settings are the same as in Table

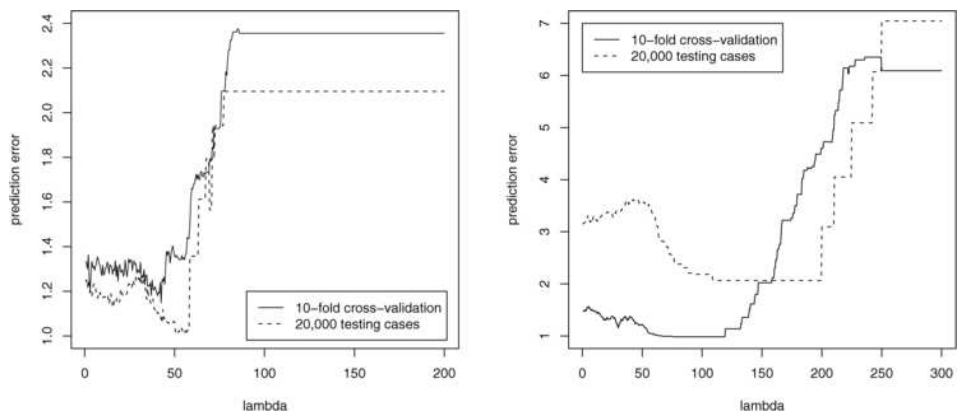

FIG. 1. *Left panel: Plot of prediction error versus $\lambda$ in $\ell_1$ regression of simulated data. Right panel: Plot of prediction error versus $\lambda$ in $\ell_2$ regression of simulated data. In both figures, the solid line represents the errors based on* 10-*fold cross-validation, and the dashed line represents the errors based on* 20,000 *testing cases.*

TABLE 2
*Simulation results of $\ell_2$ regression with a lasso penalty. Standard errors of estimates appear in parentheses. The left error column is testing error under the true parameter values; the right error column is testing error under the estimated parameter values*

| | | | | $\beta = (1, 1, 1, 1, 1, 0, \ldots, 0)$ | | | | |
|---|---|---|---|---|---|---|---|---|
| Distribution | $(p, n)$ | $\rho$ | Error | $\lambda$ | Error | $N_{\text{nonzero}}$ | $N_{\text{true}}$ | Time |
| Laplace | (5000, 200) | 0.00 | 1.97 | 112.17 | 2.13 | 5.70 | 5.00 | 0.05 |
| | | | | (16.37) | (0.15) | (2.26) | (0.00) | (0.00) |
| Laplace | (5000, 200) | 0.80 | 1.97 | 197.84 | 2.13 | 5.98 | 4.86 | 0.33 |
| | | | | (90.64) | (0.15) | (1.39) | (0.45) | (0.13) |
| Laplace | (5000, 500) | 0.00 | 2.03 | 254.62 | 2.03 | 5.20 | 5.00 | 0.11 |
| | | | | (97.20) | (0.06) | (1.13) | (0.00) | (0.01) |
| Laplace | (5000, 500) | 0.80 | 2.03 | 611.90 | 2.03 | 5.30 | 5.00 | 0.81 |
| | | | | (180.68) | (0.04) | (0.54) | (0.00) | (0.14) |
| Laplace | (50000, 500) | 0.00 | 2.02 | 255.96 | 2.03 | 5.04 | 5.00 | 0.91 |
| | | | | (71.50) | (0.05) | (0.28) | (0.00) | (0.22) |
| Laplace | (50000, 500) | 0.80 | 2.02 | 588.38 | 2.04 | 5.64 | 5.00 | 12.30 |
| | | | | (231.48) | (0.04) | (0.87) | (0.00) | (5.05) |
| Normal | (5000, 200) | 0.00 | 1.01 | 107.14 | 1.03 | 5.04 | 5.00 | 0.05 |
| | | | | (22.11) | (0.03) | (0.20) | (0.00) | (0.01) |
| Normal | (5000, 200) | 0.80 | 1.01 | 216.30 | 1.04 | 5.72 | 4.98 | 0.52 |
| | | | | (90.02) | (0.04) | (0.94) | (0.14) | (0.13) |
| Normal | (5000, 500) | 0.00 | 1.01 | 240.85 | 1.01 | 5.02 | 5.00 | 0.13 |
| | | | | (96.29) | (0.01) | (0.14) | (0.00) | (0.01) |
| Normal | (5000, 500) | 0.80 | 1.01 | 555.79 | 1.01 | 5.32 | 5.00 | 0.71 |
| | | | | (214.27) | (0.01) | (0.55) | (0.00) | (0.09) |
| Normal | (50000, 500) | 0.00 | 1.01 | 244.31 | 1.01 | 5.00 | 5.00 | 0.98 |
| | | | | (102.34) | (0.01) | (0.00) | (0.00) | (0.07) |
| Normal | (50000, 500) | 0.80 | 1.01 | 549.40 | 1.01 | 5.50 | 5.00 | 6.40 |
| | | | | (195.81) | (0.01) | (0.70) | (0.00) | (1.05) |

3 except that the residual errors follow a Laplace distribution rather than a normal distribution. The last column reports the ratio of the objective functions under the two algorithms at their converged values. Inspection of the table shows that the greedy algorithm is faster than the cyclic algorithm. Both algorithms have similar accuracy, and their accuracies are roughly comparable to the accuracies seen in Table 3 under the heading $\|\hat{\beta} - \beta\|_1$. These positive results relieve our anxieties about premature convergence with coordinate descent.

Some of the competing algorithms for $\ell_1$ regression simply do not work on the problem sizes encountered in the current comparisons. For instance, the standard iteratively reweighted least squares method proposed by Schlossmacher (1973) and Merle and Spath (1974) falters because of the large matrix inversions required. It is also hampered by infinite weights for those observations with zero residuals. We were unsuccessful in getting the standard software of Barrodale and Roberts (1980) to run properly on these large-scale problems.

TABLE 3
*Speed and accuracy of different algorithms for lasso penalized $\ell_2$ regression*

| Algorithm | $(p, n)$ | $\rho$ | $N_{\text{nonzero}}$ | $N_{\text{true}}$ | Time | $\|\hat{\boldsymbol{\beta}} - \boldsymbol{\beta}\|_1$ |
|---|---|---|---|---|---|---|
| Cyclic | (5000, 200) | 0 | 5 | 5 | 0.04 | 0.51592 |
| Greedy | | | 5 | 5 | 0.11 | 0.51567 |
| LARS | | | 94 | 5 | 2.19 | 3.33400 |
| Cyclic | (5000, 200) | 0.8 | 5 | 5 | 0.18 | 1.01544 |
| Greedy | | | 5 | 5 | 0.36 | 1.01892 |
| LARS | | | 35 | 5 | 5.45 | 1.48300 |
| Cyclic | (50000, 500) | 0 | 5 | 5 | 0.99 | 0.68995 |
| Greedy | | | 7 | 5 | 2.90 | 0.68700 |
| LARS | | | | not available | | |
| Cyclic | (50000, 500) | 0.8 | 5 | 5 | 4.11 | 0.60956 |
| Greedy | | | 5 | 5 | 7.94 | 0.60875 |
| LARS | | | | not available | | |
| Cyclic | (500, 5000) | 0 | 5 | 5 | 0.24 | 0.06338 |
| Greedy | | | 5 | 5 | 0.36 | 0.06336 |
| LARS | | | 27 | 5 | 0.78 | 0.25370 |
| Cyclic | (500, 5000) | 0.8 | 5 | 5 | 0.30 | 0.11082 |
| Greedy | | | 5 | 5 | 0.71 | 0.11049 |
| LARS | | | 14 | 5 | 1.168 | 0.18884 |

**8. Analysis of real data.** We now turn to real data involving gene expression levels and obesity in mice. Wang et al. (2006b) measured abdominal fat mass on $n = 311$ F2 mice (155 males and 156 females). The F2 mice were created by mating two inbred strains and then mating brother-sister pairs from the resulting offspring. Wang et al. also recorded the expression levels in liver of $p = 23{,}388$ genes in each mouse.

TABLE 4
*Comparison of greedy and cyclic coordinate descent for lasso penalized $\ell_1$ regression*

| Algorithm | $(p, n)$ | $\rho$ | $N_{\text{nonzero}}$ | $N_{\text{true}}$ | Time | $\|\hat{\boldsymbol{\beta}} - \boldsymbol{\beta}\|_1$ | $\frac{f_{\text{greedy}}}{f_{\text{cyclic}}}$ |
|---|---|---|---|---|---|---|---|
| Greedy | (5000, 200) | 0 | 7 | 5 | 0.02 | 0.84228 | 1.01063 |
| Cyclic | | | 7 | 5 | 0.10 | 0.91861 | ($\lambda = 50$) |
| Greedy | (5000, 200) | 0.8 | 5 | 5 | 0.04 | 0.53354 | 0.99118 |
| Cyclic | | | 6 | 5 | 0.39 | 0.74330 | ($\lambda = 57.58$) |
| Greedy | (50000, 500) | 0 | 5 | 5 | 0.34 | 0.32212 | 1.00288 |
| Cyclic | | | 5 | 5 | 3.99 | 0.28565 | ($\lambda = 124.5$) |
| Greedy | (50000, 500) | 0.8 | 7 | 5 | 1.66 | 0.97379 | 1.00018 |
| Cyclic | | | 8 | 5 | 8.66 | 0.84372 | ($\lambda = 110$) |
| Greedy | (500, 5000) | 0 | 5 | 5 | 0.07 | 0.06680 | 0.99938 |
| Cyclic | | | 5 | 5 | 1.01 | 0.06604 | ($\lambda = 1144.14$) |
| Greedy | (500, 5000) | 0.8 | 5 | 5 | 0.13 | 0.12882 | 1.00008 |
| Cyclic | | | 5 | 5 | 1.53 | 0.12943 | ($\lambda = 1144.14$) |

Our first model postulates that the fat mass $y_i$ of mouse $i$ satisfies

$$y_i = 1_{\{i \text{ male}\}}\mu_1 + 1_{\{i \text{ female}\}}\mu_2 + \sum_{j=1}^{p} x_{ij}\beta_j + \epsilon_i,$$

where $x_{ij}$ is the expression level of gene $j$ of mouse $i$ and $\epsilon_i$ is random error. Since male and female mice exhibit across the board differences in size and physiology, it is prudent to estimate a different intercept for each sex. The left panel of Figure 2 plots as a function of $\lambda$ the average number of nonzero predictors and the average prediction error. Here we use $\ell_1$ regression and 10-fold cross-validation. The right panel of Figure 2 plots the same quantities under $\ell_2$ regression.

The 10-fold cross-validation curve $c(\lambda)$ is ragged under both $\ell_1$ and $\ell_2$ regression. For $\ell_1$ regression, examination of $c(\lambda)$ over a fairly dense grid shows an optimal $\lambda$ of about 3.5. Here the average number of nonzero predictors is 88.5, and the average testing error is 0.6533. For the entire data set, the number of nonzero predictors is 77, and the training error is 0.4248. For $\ell_2$ regression, the optimal $\lambda$ is 7.8. Here the average numbers of predictors is 36.8, and the average testing error is 0.7704. For the entire data set, the number of nonzero predictors is 41, and the training error is 0.3714. The preferred $\ell_1$ and $\ell_2$ models share 27 predictors in common.

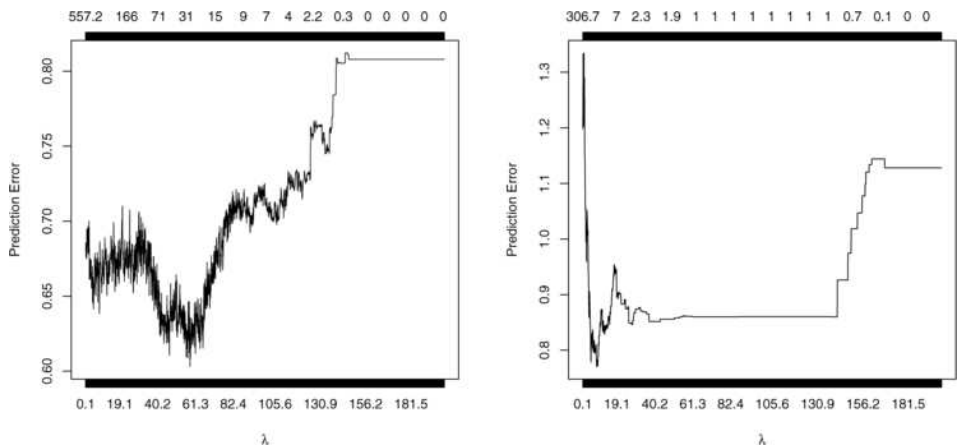

FIG. 2. *Left panel: Plot of 10-fold cross-validation error and number of predictors versus $\lambda$ in $\ell_1$ regression of the mice microarray data. Right panel: Plot of 10-fold cross-validation error and number of predictors versus $\lambda$ in $\ell_2$ regression of the mice microarray data. The lower x-axis plots the values of $\lambda$, and the upper x-axis plots the number of predictors. The y-axis is prediction error based on 10-fold cross-validation versus $\lambda$.*

Given the inherent differences between the sexes, it is enlightening to assign sex-specific effects to each gene and to group parameters accordingly. These decisions translate into the model

$$y_i = 1_{\{i \text{ male}\}} \left(\mu_1 + \sum_{j=1}^{p} x_{ij}\beta_{1j}\right) + 1_{\{i \text{ female}\}} \left(\mu_2 + \sum_{j=1}^{p} x_{ij}\beta_{2j}\right) + \epsilon_i$$

$$= \sum_{j=0}^{p} z_{ij}^t \gamma_j + \epsilon_i,$$

where the bivariate vectors $z_{ij}$ and $\gamma_j$ group predictors and parameters, respectively. Thus,

$$z_{ij}^t = \begin{cases} (1,0), & i \text{ is male and } j=0, \\ (0,1), & i \text{ is female and } j=0, \\ (x_{ij},0), & i \text{ is male and } j>0, \\ (0,x_{ij}), & i \text{ is female and } j>0, \end{cases}$$

and $\gamma_0^t = (\mu_1, \mu_2)$ and $\gamma_j^t = (\beta_{1j}, \beta_{2j})$ for $j > 0$. In this notation, the objective function becomes

$$f(\theta) = \tfrac{1}{2}\sum_{i=1}^{n}\left(y_i - \sum_{j=0}^{p} z_{ij}^t \gamma_j\right)^2 + \lambda_2 \sum_{j=1}^{q}\|\gamma_j\|_2 + \lambda_1 \sum_{j=1}^{q}\|\gamma_j\|_1.$$

Under 10-fold cross-validation, the optimal pair $(\lambda_1, \lambda_2)$ occurs at approximately $(5,1)$. This choice leads to an average of 40.1 nonzero predictors and an average prediction error of 0.8167. For the entire data set, the number of nonzero predictors is 44, and the training error is 0.3128. Among the 44 predictors, three are paired female–male slopes. Thus, the preferred model retains 41 genes in all. Among these 41 genes, 25 appear in the $\ell_1$ model, and 26 appear in the $\ell_2$ model without the group penalties. For all three models, there are 20 genes in common.

In carrying out these calculations, we departed from the tack taken by Wang et al. (2006b), who used marker genotypes rather than expression levels as predictors. Our serendipitous choice identified some genes known to be involved in fat metabolism and turned up some interesting candidate genes. One of the known genes from the short list of 20 just mentioned is pyruvate dehydrogenase kinase isozyme 4 (Pdk4). This mitochondrial enzyme, which has been studied for its role in insulin resistance and diabetes, is a negative regulator of the pyruvate dehydrogenase complex. The upregulation of Pdk4 promotes gluconeogenesis. In the coexpression network studies of Ghazalpour et al. (2005, 2006), Pdk4 was one of the genes found in the module associated with mouse body weight. Here we find that expression of Pdk4 is a good predictor of fat pad mass. It has been suggested that enhanced PDK4 expression is a compensatory mechanism countering the excessive formation of intracellular lipid and the exacerbation of impaired insulin sensitivity [Huang et al. (2002) and Sugden (2003)].

A second gene on our list is leukotriene a4 hydrolase. This enzyme converts leukotriene A4 to leukotriene B4 as part of the 5-lipooxygenase inflammatory pathway, which Mehrabian et al. (2005) report influences adiposity in rodents. Other genes on our list include three involved in energy metabolism: Ckm, which plays a central role in energy transduction; 3-hydroxyisobutyrate dehydrogenase, which is part of the valine, leucine and isoleucine degradation pathway previously reported to be associated with subcutaneous fat pad mass in a different mouse cross [Ghazalpour et al. (2005)]; and the thiamine metabolism pathway gene Nsf1, which uses an endproduct of the steroid metabolism pathway. In addition, we identified several genes with no obvious ties to fat pad mass, energy metabolism, or obesity in general, including three riken cDNAs, Plekha8, Gloxd1, the signaling molecule Rras, and the transcription factor Foxj3. All of these are also present in our larger list of 27 genes ignoring sex dependent slopes. This larger list includes another transcription factor, an olfactory receptor gene, and an adhesion molecule. The olfactory receptor gene is particularly intriguing because it could affect feeding behavior in mice.

**9. Discussion.** Lasso penalized regression performs continuous model selection by estimation rather than by hypothesis testing. Several factors converge to make penalized regression an ideal exploratory data analysis tool. One is the avoidance of the knotty issues of multiple testing. Another is the sheer speed of the coordinate descent algorithms. These algorithms offer decisive advantages in dealing with modern data sets where predictors wildly outnumber cases. If Fu (1998) had written his paper a few years later, this trend would have been clearer, and doubtless he would not have concentrated on small problems with cases outnumbering predictors.

We would not have predicted beforehand that the normally plodding coordinate descent methods would be so fast. In retrospect, it is clear that their avoidance of matrix operations and quick dismissal of poor predictors make all the difference. Our initial concern about the adequacy of Edgeworth's algorithm have been largely laid to rest by the empirical evidence. On data sets with an adequate number of cases, the poor behavior reported in the past does not predominate for either version of coordinate descent.

Although the supplementary appendix [Wu and Lange (2008)] proves that lasso constrained $\ell_1$ regression is consistent, parameter estimates are biased toward zero in small samples. For this reason, once we have identified the active parameters for a given value of the tuning constant $\lambda$, we re-estimate them ignoring the lasso penalty. Failure to make this adjustment tends to favor smaller values of $\lambda$ in cross-validation and the inclusion of irrelevant predictors in the preferred model.

Better understanding of the convergence properties of the algorithms is sorely needed. Tseng (2001) proves convergence of cyclic coordinate descent for $\ell_2$ penalized regression. Our treatment of greedy coordinate descent in the Appendix is different and simpler. Neither proof determines the rate of convergence. Even more pressing is the challenge of overcoming the theoretical defects of Edgeworth's algorithm without compromising its speed. On a practical level, the reparameterization of Li and Arce (2004), which operates on pairs of parameters, may allow Edgeworth's algorithm to escape many trap points. If this tactic is limited to the active parameters, then speed may not degrade unacceptably.

Our algorithm for grouped parameters exploits a majorization used in constructing other MM algorithms. The techniques and theory behind MM algorithms deserve to be better known [Hunter and Lange (2004)]. Majorization approximately doubles the number of iterations until convergence in the $\ell_2$ cyclic coordinate descent algorithm. Although both the original and the grouped $\ell_2$ algorithms take hundreds of iterations to converge, each iteration is so cheap that overall speed is still impressive.

Lasso penalized estimation extends far beyond regression. The papers of Fu (1998) and Park and Hastie (2006a, 2006b) discuss some of the possibilities in generalized linear models. We have begun experimenting with cyclic coordinate descent in logistic regression. Although explicit maxima for the one-dimensional subproblems are not available, Newton's method converges reliably in a handful of steps. The results are very promising and will be dealt with in another paper.

## APPENDIX: CONVERGENCE THEORY

Our proof of Proposition 1 splits into a sequence of steps. We first show that a minimum exists. This is a consequence of the continuity of $f(\theta)$ and the coerciveness property that $\lim_{\|\theta\|_2 \to \infty} f(\theta) = \infty$. If any $|\beta_j|$ tends to $\infty$, then the claimed limit is obvious. If all $\beta_j$ remain bounded but $|\mu|$ tends to $\infty$, then each squared residual $(y_i - \mu - x_i^t \beta)^2$ tends to $\infty$.

Selection of which component of $f(\theta)$ to update is governed by

$$h(\theta) = \min_i \min\{d_{e_i} f(\theta), d_{-e_i} f(\theta)\}.$$

Although the function $h(\theta)$ is not continuous, it is upper semicontinuous. This weaker property will be enough for our purposes. Upper semicontinuity means that $\limsup_{m \to \infty} h(\theta^m) \le h(\theta)$ whenever $\theta^m$ converges to $\theta$ [Rudin (1987)]. The collection of upper semicontinuous functions includes all continuous functions and is closed under the formation of finite sums and minima.

Every directional derivative of the residual sum of squares is continuous. Thus, to verify that $h(\theta)$ is upper semicontinuous, it suffices to show that the coordinate directional derivatives of the penalty terms $\lambda|\beta_j|$ are upper semicontinuous. In a small enough neighborhood of $\beta_j \neq 0$, all coordinate directional derivatives of $\lambda|\beta_j|$ are constant and therefore continuous. At $\beta_j = 0$ the directional derivatives along $e_j$ and $-e_j$ are both $\lambda$, the maximum value possible. Hence, the limiting inequality of upper semicontinuity holds.

Any discussion of convergence must take into account the stationary points of the algorithm. Such a point $\theta$ satisfies the conditions $d_{e_j} f(\theta) \geq 0$ and $d_{-e_j} f(\theta) \geq 0$ for all $j$. If we let $\mu$ vary along the coordinate direction $e_0$, then straightforward calculations produce the general directional derivative

$$d_v f(\theta) = \sum_j \frac{\partial}{\partial \theta_j} g(\theta) v_j + \lambda \sum_{j>0} \begin{cases} v_j, & \theta_j > 0, \\ -v_j, & \theta_j < 0, \\ |v_j|, & \theta_j = 0. \end{cases}$$

It follows that

$$d_v f(\theta) = \sum_{v_j > 0} d_{e_j} f(\theta) v_j + \sum_{v_j < 0} d_{-e_j} f(\theta) |v_j|$$

and that every directional derivative is nonnegative at a stationary point.

Because $f(\theta)$ is convex, the difference quotient $s^{-1}[f(\theta + sv) - f(\theta)]$ is increasing in $s > 0$. Therefore, $f(\theta + v) - f(\theta) \geq d_v f(\theta)$, and if $\theta$ is a stationary point, then $f(\theta + v) \geq f(\theta)$ for all $v$. In other words, $\theta$ is a minimum point. Conversely, it is trivial to check that $d_v f(\theta) \geq 0$ for every $v$ when $\theta$ is a minimum point. Hence, stationary points and minimum points coincide.

With these preliminaries in mind, suppose the sequence $\theta^m$ generated by greedy coordinate descent has a subsequence $\theta^{m_k}$ converging to a nonstationary point $\theta^*$. By virtue of semicontinuity, we have

(A.1) $$h(\theta^{m_k}) \leq \tfrac{1}{2} h(\theta^*) < 0$$

for infinitely many $k$. We will demonstrate that this inequality forces the decreasing sequence $f(\theta^m)$ to converge to $-\infty$, contradicting the fact that $f(\theta)$ is bounded below. In fact, we will demonstrate that there exists a constant $c > 0$ with $f(\theta^{m_k+1}) \leq f(\theta^{m_k}) - c$ for all $\theta^{m_k}$ satisfying inequality (A.1). The existence of the constant $c$ is tied to the second derivatives

$$\frac{\partial^2}{\partial \theta_j^2} f(\theta) = \sum_{i=1}^n x_{ij}^2$$

of $f(\theta)$ along each coordinate direction. Let $b = \max_j \sum_{i=1}^n x_{ij}^2$.

Now suppose that $\theta^{m_k}$ satisfies inequality (A.1). For notational simplicity, let $y = \theta_j$ be the component being updated, $x = \theta_j^{m_k}$, and $s(y) = f(\theta)$ as a function of $\theta_j$. Provided we restrict $y$ to the side of 0 showing the most negative directional derivative, $s(y)$ is twice differentiable and satisfies the majorization

$$s(y) = s(x) + s'(x)(y-x) + \tfrac{1}{2}s''(z)(y-x)^2$$
$$\leq s(x) + s'(x)(y-x) + \tfrac{1}{2}b(y-x)^2.$$

If $w = x - \frac{s'(x)}{b}$ denotes the minimum of the majorizing quadratic, then

$$s(w) \leq s(x) + s'(x)(w-x) + \frac{1}{2}b(w-x)^2 \;=\; s(x) - \frac{1}{2}\frac{s'(x)^2}{b}.$$

At the minimum $z$ of $s(y)$ we have

$$s(z) \leq s(w) \leq s(x) - \frac{1}{2}\frac{s'(x)^2}{b}.$$

This identifies the constant $c$ as $c = \frac{1}{2b}[\frac{1}{2}h(\theta^*)]^2$ and completes the proof of the proposition.

**Acknowledgments.** We thank the Department of Statistics at Stanford University for hosting our extended visit during the 2006–2007 academic year. Jake Lusis of UCLA kindly permitted us to use the Wang et al. (2006b) mouse data. Anatole Ghazalpour helped interpret the statistical analysis of these data. Emanuel Candes pointed out references pertinent to the history of the lasso and clarified our thinking about convergence of the $\ell_2$ greedy algorithm.

## SUPPLEMENTARY MATERIAL

**Proof of weak consistency of Lasso penalized $\ell_1$ regression**
(doi: 10.1214/07-AOAS147SUPP; .pdf). Our supplementary appendix demonstrates the weak consistency of penalized $\ell_1$ estimators. The proof is a straightforward adaptation of the arguments of Oberhofer (1983) on nonlinear $\ell_1$ regression. Since we only consider linear models, the regularity conditions in Oberhofer (1983) are relaxed and clarified.

| | |
|---|---|
| Department of Epidemiology and Biostatistics | Department of Biomathematics, Human Genetics and Statistics |
| University of Maryland | University of California |
| College Park, Maryland 20707 | Los Angeles, California 90095 |
| USA | USA |
| E-mail: ttwu@umd.edu | E-mail: klange@ucla.edu |